\documentclass{moriond}

\def\Journal#1#2#3#4{{#1} {\bf #2}  (#4) #3}

\def\PRL{\em Phys. Rev. Lett.}
\def\PRD{{\em Phys. Rev.} D}

\def\be{\begin{equation}}
\def\ee{\end{equation}}
\def\bea{\begin{Eqarray}}
\def\eea{\end{Eqarray}}

\newcommand{\Photo}{}

\usepackage{amsmath}
\usepackage{slashed}
\usepackage{tikz-feynman}

\begin{document}
	
{\flushright
DO-TH 19/07

QFET-2019-05

}

\vspace*{3cm}
\title{ASYMPTOTICALLY SAFE EXTENSIONS OF THE STANDARD MODEL WITH FLAVOUR PHENOMENOLOGY}

\author{Gudrun Hiller${}^1$, Clara Hormigos-Feliu${}^1$, Daniel F. Litim${}^2$ and  Tom Steudtner${}^2$}

\address{${}^1$Fakult\"at Physik, TU Dortmund, Otto-Hahn-Str.4, 
	D-44221 Dortmund, Germany\\
	${}^2$Department of Physics and Astronomy, University of Sussex, Brighton,
	 BN1 9QH, U.K.}

\maketitle\abstracts{ 
We investigate  asymptotically safe extensions of the Standard Model with new matter fields arising in the TeV energy range.  The new sector contains singlet scalars and vector-like fermions in representations which permit Yukawa interactions with the Standard Model leptons. Phenomenological implications are explored including charged lepton flavour violation, Drell-Yan processes and lepton anomalous magnetic moments.  For the latter, we find that BSM contributions can be sizeable enough to explain the present experimental discrepancies of the electron and muon anomalous magnetic moments with the Standard Model.
}
\section{Asymptotically safe models}

In particle physics,  the running of couplings is a consequence of quantum fluctuations. In the Standard Model (SM), it causes the strong and weak
gauge couplings to become asymptotically free while the 
hypercharge coupling, ultimately, runs into a Landau pole. Asymptotic safety (AS) generalizes the paradigm of asymptotic freedom by observing that couplings may very well run into interacting rather than  free fixed points at high energies.\cite{guaranteed,theorems}
The motivation to search for  asymptotically safe extensions of the SM, then, is twofold. On a fundamental level,  asymptotically safe theories  are UV complete and predictive at energies up to, at least, the Planck scale. On the phenomenological side, asymptotic safety opens a door into uncharted territory for model building.  Moreover, an extra benefit is that the predictivity of asymptotically safe models is often enhanced, in the sense that the existence of  (weakly) interacting fixed points and UV--IR connecting renormalisation group trajectories  put tight constraints on models and parameters such as couplings or masses of BSM matter fields.\cite{directions}

On a technical level, asymptotic safety relates to an interacting UV  fixed point $(\alpha^*\ge 0)$  which are the zeros of the renormalisation group $\beta$-functions of all couplings $\alpha$, schematically written as
\begin{equation}\label{0}
\left. \beta_\alpha \right\vert_{*} \equiv \left. \frac{d \alpha}{d \ln \mu}\right\vert_{\alpha=\alpha^*} = 0 \,.
\end{equation}
At weak coupling, the necessary and sufficient conditions for interacting fixed points  (\ref{0}) are well-understood.\cite{theorems} Most notably, Yukawa couplings are essential to negotiate weakly interacting UV fixed points.\cite{guaranteed,theorems} Since asymptotic safety is not realised in the  SM as we know it today, this implies that in order to render an extension of the SM asymptotically safe, we need to introduce BSM matter fields and at least one, if not several, new Yukawa couplings.
A setup which consistently gives rise to weakly interacting UV fixed points in simple and semi-simple gauge theories contains $N_F$ vector-like fermions $\psi_i$, with charges under an $SU(N)$  gauge group, and $N_F^2$ complex singlet scalars $S_{ij}$.\cite{guaranteed,theorems} First explorations of asymptotically safe  SM extensions with a  Yukawa sector $\sim y {\rm Tr}\,[ \overline{\psi}_L S \psi_R]$ and their phenomenology have also been studied.\cite{directions}

In this work, we extend the range of asymptotically safe models by introducing  Yukawa couplings between SM and BSM matter fields, which thereby  serve as a portal to new physics.\cite{paper} The main novelty is that these models become sensitive to flavour physics.
This idea can be implemented in a minimal way by choosing suitable representations for the BSM fermions  $\psi$ in order to permit renormalisable Yukawa couplings with the SM fermions. From now on,
we focus on the case where the $\psi$ are singlets under $\rm SU(3)_C\times SU(2)_L$ and carry a hypercharge $Y_F = -1$; a full analysis of further  models is reported elsewhere.\cite{paper}  We  also take $N_F=3$, which enables us to  construct interactions with the SM in flavour space. In what follows, we refer to this specific scenario as model A. The BSM Yukawa Lagrangian in this model reads
\begin{equation}\label{modelA}
-\mathcal{L}_Y^{\rm A} =  y\, {\rm Tr}\,[ \overline{\psi}_L S \psi_R] + \kappa\,\overline{L}H\psi_R +  \kappa'\, {\rm Tr} [\overline{E} S^{\dagger}\psi_L] +  h.c.  ,
\end{equation}
where $L,E$ are the SM lepton doublet and singlet respectively,  $H$ is the Higgs doublet, and the traces are taken over all gauge and flavour indices. The fact that the couplings $\kappa$ and $\kappa'$ may contain a nontrivial flavour structure is taken into account when discussing their phenomenological signatures below. For renormalisation group evolution purposes, however, we consider all Yukawa couplings to be  real parameters. 

We have derived the renormalisation group equations for the model with \eqref{modelA} and its fixed points \eqref{0} up to the complete second order in the loop expansion.  Fixed points for all couplings come out moderately small, or zero. For example,  in  \eqref{modelA}, the ``reduced'' Yukawa couplings   in units of the perturbative loop factor ($e.g.$~$\alpha_{\kappa}\equiv \kappa^2/(16\pi^2)$, and similarly for  other couplings)  are typically of order ${\cal O}(0.1)$ in the deep UV. 
We have also investigated trajectories which connect the asymptotically safe high energy regime with the SM at low energies. We find that both the Higgs  and the BSM scalar sector have a stable ground state for all scales.
On the other hand, not all UV-safe trajectories can be matched to the SM, but if they can, they impose constraints for the BSM parameters at the matching scale.
For example, we find that a matching to the SM at the scale $M_F=1$ TeV can be achieved, with $\alpha_{\kappa},\alpha_{\kappa'}\simeq 3\cdot 10^{-3}$. Remarkably, the Yukawa coupling $y$ is not required and may vanish at all scales because it corresponds to an irrelevant coupling in the UV. We conclude that the $\kappa$ and $\kappa'$ interactions in \eqref{modelA} are sufficient to deliver an asymptotically safe scenario which can be connected to the SM in the TeV energy range.

\section{Phenomenology and anomalous magnetic moments}

\begin{figure}[h]
	\centering
	\begin{minipage}{0.52\linewidth}
		\centerline{\includegraphics[width=0.79\linewidth]{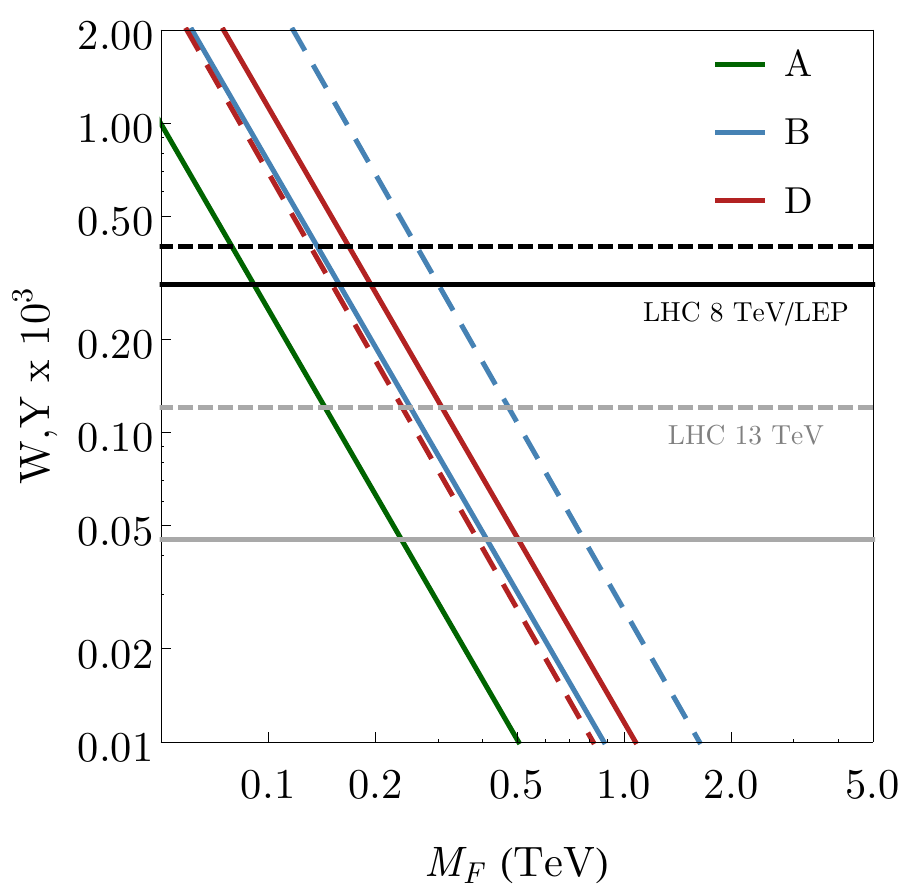}}
		\caption[]{Present (black) and future (grey) constraints from EW precision tests \cite{rudermann}. In addition to model A where the $\psi$ carry SM charges $({\bf 1},{\bf 1}, -1)$, we also show constraints for models B and D where the $\psi$ carry SM
		charges $({\bf 1},{\bf 3}, -1)$ and $({\bf 1},{\bf 2},-3/2)$ respectively.}
		\label{fig:WY}
	\end{minipage}
	\hfill
	\begin{minipage}{0.45\linewidth}
			\vspace{-0.15cm}
		\centerline{\includegraphics[width=0.99\linewidth]{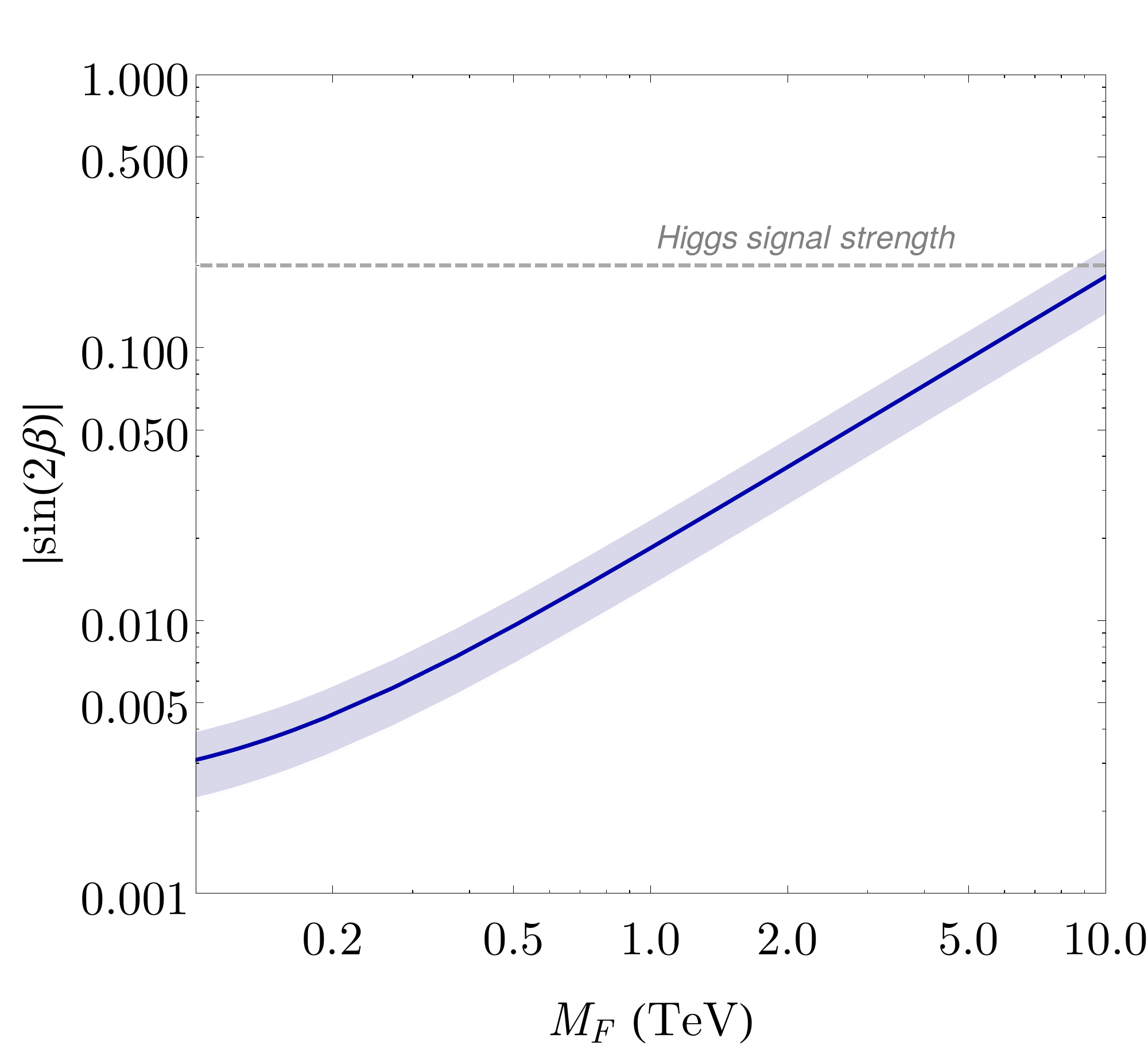}}
		\caption[]{
			Values of $M_F$ and mixing angle $\beta$ yielding a contribution equal to $\Delta a_{\mu}$ within a $1\sigma$ range. The parameter space corresponds to the contribution~\eqref{g-2en}, which presents a chiral flip on the heavy fermion line. }
		\label{fig:beta}
	\end{minipage}
\end{figure}

The new heavy fields can be produced at hadron and lepton colliders either through gauge interactions (in the case of the $\psi$) or through the $\kappa$ and $\kappa'$ couplings (both $\psi$ and $S$). Drell-Yan processes in particular provide the opportunity to study their effects through electroweak precision tests,  specifically in terms of the parameters $W$ and $Y$. These are sensitive to $M_F$ and to the deviations $\Delta B$ of the one-loop coefficients in the $\beta$-functions of the gauge couplings, scaling as 
\begin{equation}
W,Y \propto \,\alpha_{1,2}\,\Delta B_{1,2}\frac{M_W^2}{M_F^2}\,.
\end{equation}
We find that present bounds \cite{rudermann} constrain $M_F$ to be at least 0.1 TeV for model A with  (\ref{modelA}),  and heavier for other models with similar features, as shown in Fig.~\ref{fig:WY}.

\begin{figure}
	\centering
	\begin{minipage}{0.48\linewidth}
		\centering
		\centerline{\includegraphics[width=0.63\linewidth]{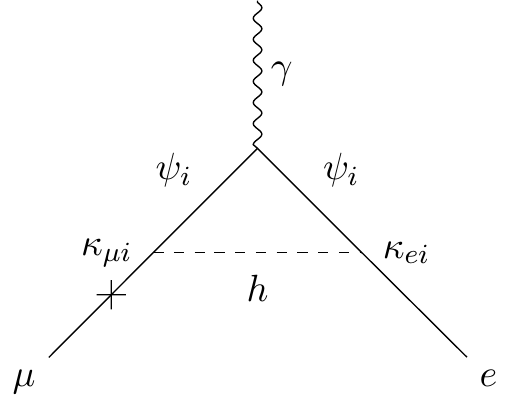}}
		\caption[]{
			Contribution to $\mu\to e\gamma$ involving off-diagonal elements of $\kappa$, where $h$ is the SM Higgs. The chiral flip ($\times$) is on the line of the decaying lepton.}
		\label{fig:LFV}
	\end{minipage}
\hfill
	\begin{minipage}{0.48\linewidth}
	\centering
		\centerline{\includegraphics[width=0.63\linewidth]{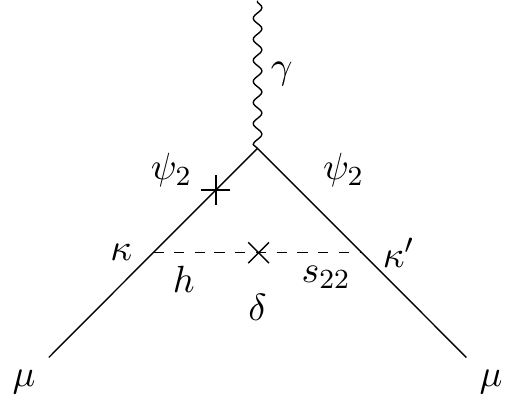}}
	\caption[]{Contribution to $a_{\mu}$ with a chiral flip on the heavy fermion line, where $s_{22}$ is the component of $S$ which mixes with $h$.}
	\label{fig:g-2}
\end{minipage}
\end{figure}

On the other hand, charged lepton flavour violation (cLFV) is induced by the BSM Yukawas $\kappa$ and $\kappa'$ provided they contain some non-vanishing off-diagonal elements. Flavour-violating decays of leptons as well as $\mu-e$ conversion in nuclei are then induced at one-loop, and thus constitute sensitive probes of the flavour structure of the BSM sector. Radiative processes involve diagrams which require a chiral flip in a fermion line, as seen in Fig.~\ref{fig:LFV} for $\mu\to e \gamma$. We find that present bounds allow $\kappa_{\mu i}\kappa_{ei}/16\pi^2\sim 10^{-4}$ or smaller, while processes involving $\tau$ leptons remain less constrained, with $\kappa_{\tau i}\kappa_{ei, \mu i}/16\pi^2$ up to $\mathcal{O}(10^{-1})$ still allowed.\cite{paper}

In model A, the anomalous magnetic moments of the leptons receive contributions through scalar exchange (similarly to Fig.~\ref{fig:LFV}, with same-fermion external legs). These scale as $a_\ell^{\rm NP}\sim \alpha_{\kappa}\,m_\ell^2/M_F^2$, where $a_\ell$ and $m_\ell$ are the anomalous magnetic moment and mass of the lepton $\ell$, respectively. In the case of the muon, the long-standing discrepancy between measurements and theory predictions amounts to $\Delta a_{\mu} = 268(63)(43) \cdot 10^{-11}$, which represents a $3.5\,\sigma$ deviation from the SM.\cite{g-2}  For the electron, recent measurements of  $\Delta a_{e} = - 88(28)(23) \cdot 10^{-14}$ lead to a pull of $-2.3\,\sigma$  from the SM prediction.\cite{g-2e} The aforementioned contributions in model A add up to less than 1\% of these anomalies for typical values  of $\alpha_\kappa$ of the size of weak couplings. On the other hand, if the scalar sector presents a non-vanishing quartic coupling $\delta H^{\dagger}H {\rm Tr} [S^{\dagger} S]$ a different contribution is possible which presents a chiral flip in a heavy fermion line (see Fig.~\ref{fig:g-2}), and can therefore be potentially enhanced with respect to the contributions discussed above. This mechanism requires at least one component of $S$ to acquire a vacuum expectation value (VEV), which leads to $H-S$ mixing through the portal $\delta$. 

The BSM scalar potential of the model allows for two non-trivial vacuum configurations. We focus now on the configuration  $V^-$, which presents a VEV in only one of the diagonal components of $S$. The scenario of $V^-$ has flavour-specific consequences: If the $S_{22}$ component is chosen to take a VEV, the second-generation fermions are singled out, since only they interact with the two mixing scalars. Then, in the Lagrangian~\eqref{modelA} the Yukawas $\kappa$ and $\kappa'$ allow for couplings of left- and right-handed muons with both new scalar mass eigenstates, which makes the diagram in Fig.~\ref{fig:g-2} possible. In this scenario, the dominant new physics contribution to the anomalous magnetic moment of the muon reads
\begin{equation}\label{g-2en}
 a_{\mu}^{\rm NP}  =  -\frac{m_{\mu}}{2 M_F} \frac{\kappa\kappa'}{16\pi^2} \sin 2\beta  \propto \delta \frac{m_h}{M_S}\frac{m_\mu}{M_F}\frac{\kappa\kappa'}{16\pi^2}\,,
\end{equation}
where $M_S$ is the mass of the $S$ and $\beta$ is the scalar mixing angle. If $\kappa$ and $\kappa'$ are of the same order of magnitude, as found in benchmarks of model $A$ where matching is possible, the contribution~\eqref{g-2en} presents an enhancement by a factor $\sim M_F/m_{\mu}\sin 2\beta$ with respect to $a_{\mu}^{\rm NP}$ in the absence of portal coupling, owing to the shift of the chiral flip from the muon to the $\psi$ line. The parameter space yielding a contribution equal to $\Delta a_{\mu}$ is shown in Fig.~\ref{fig:beta}. For $M_F=1$ TeV, a value of $\lvert \beta\rvert\sim0.01$ accommodates experimental observations. This is consistent with Higgs signal strength measurements, which give an order of magnitude weaker bounds.\cite{g-2} Furthermore, cLFV processes involving muons acquire in this case an enhancement analogous to that of $a_{\mu}$, with branching ratios being affected by an additional factor proportional to $M_F^2/m_{\mu}^2 \sin^2 2\beta$. Thus, in the muon-specific $V^-$ scenario cLFV constraints only allow $\kappa_{\mu i}\kappa_{ei}/16\pi^2$ up to $\mathcal{O}(10^{-9})$.
In the other possible vacuum structure, $V^+$, all diagonal components of $S$ acquire a universal VEV. Hence, resolving the tension in both $a_e$ and $a_{\mu}$ is in this case possible. It requires, however, more flavour structure in the BSM Yukawa sector. The present values of $\Delta a_{\mu}$ and $\Delta a_e$ can be explained if one invokes a suppression in the electron coupling by $\kappa = {\rm diag}(\epsilon\kappa,\kappa,\kappa)$, with $\epsilon \sim -0.05$. In the $V^+$ scenario, all cLFV decays of a lepton $\ell$ pick up an extra factor $M_F^2/m_{\ell}^2 \sin^2 2\beta$, and are as well sensitive to the possible flavour structure of $\kappa$ and $\kappa'$.

\section{Conclusions}

We considered extensions of the SM which becomes asymptotically safe owing to new Yukawa interactions between vector-like fermions and leptons. The effects of the BSM sector can be probed through electroweak precision tests, charged lepton flavour violation and anomalous magnetic moments of the leptons. Regarding the latter, we find that the presence of the BSM couplings $\kappa$ and $\kappa'$ in our model \eqref{modelA}, together with mixing in the scalar sector, give rise to chirally enhanced contributions to the anomalous magnetic moments of the muon and the electron which explain the present tensions with the SM. Other features, and additional models with alternative Yukawa sectors, are discussed  elsewhere.\cite{paper} We conclude that asymptotically safe extensions of the SM can provide predictive scenarios with relevant phenomenological implications.

\section*{Acknowledgements}

CHF is indebted to the organizers for the opportunity to present this work and for the financial support. This work has been supported by the DFG Research Unit FOR 1873 "Quark Flavour Physics and Effective Field Theories".


\section*{References}

\begin{thebibliography}{99}

\bibitem{guaranteed} 
DF~Litim and F~Sannino, \Journal{\em JHEP}{12}{178}{2014},
DF~Litim, M~Mojaza, and F~Sannino, \Journal{\em JHEP}{01}{081}{2016},
AD Bond, DF Litim, G Medina Vazquez, and T Steudtner, \Journal{\PRD}{97}{036019}{2018}.
\bibitem{theorems} AD Bond and DF Litim, \Journal{\em Eur. Phys. J.}{C77}{429}{2017}, 
\Journal{\PRL}{119}{211601}{2017}, 
\Journal{\PRD}{97}{085008}{2018},
 arXiv:1801.08527 (PRL, to appear). 

\bibitem{directions} AD Bond, G Hiller, K Kowalska, and DF Litim, \Journal{\em JHEP}{08}{004}{2017},
K Kowalska, AD Bond, G. Hiller, and DF Litim, 
\Journal{\em PoS}{EPS-HEP2017}{542}{2017}.

\bibitem{paper} G Hiller, C Hormigos-Feliu, DF Litim and T Steudtner, to appear.

\bibitem{rudermann} DSM Alves, J Galloway, JT Ruderman, and JR Walsh, \Journal{\em JHEP}{02}{007}{2015}.

\bibitem{g-2} M Tanabashi {\em et al}, \Journal{\PRD}{98}{03001}{2018}.

\bibitem{g-2e} RH Parker, C Yu, W Zhong, B Estey, and H M\"uller, \Journal{\em Science}{360}{191}{2018}.


\end{thebibliography}
\end{document}